\documentclass[aip, amsmath, amssymb,reprint]{revtex4-1}    
\usepackage{bm}
\usepackage{graphicx}
\usepackage[utf8]{inputenc}
\usepackage[T1]{fontenc}
\usepackage{mathptmx}
\usepackage{hhline}

\def \d{\partial}
\newcommand{\dd}{\partial}
\def \bv{{\bf v}}

\def \br{{\bf r}}

\def \bB{{\bf B}}

\def \bR{{\bf R}}
\def \br{{\bf r}}

\def \bk{{\bf k}}

\def \varkappa{\kappa}

\begin{document}

\title{Evolution of localized magnetic field perturbations and the nature of turbulent dynamo.
}

\author{ A.S. Il'yn}
\altaffiliation[Also at ]{National Research University Higher School of Economics, 101000,
Myasnitskaya 20, Moscow, Russia}
\email{asil72@mail.ru}
\author{ A.V. Kopyev}
\email{kopyev@lpi.ru}
\author{V.A. Sirota}
\email{sirota@lpi.ru}
\author{K.P. Zybin}
\altaffiliation[Also at ]{National Research University Higher School of Economics, 101000,
Myasnitskaya 20, Moscow, Russia}
\email{zybin@lpi.ru}
\affiliation{P.N.Lebedev Physical Institute of RAS, 119991, Leninskij pr.53, Moscow,
Russia}


\begin{abstract}

Kinematic dynamo in incompressible isotropic  turbulent flows with high magnetic Prandtl number is considered. The approach interpreting 
an arbitrary magnetic field distribution as a superposition of localized perturbations (blobs) is proposed. We derive a relation between
 stochastic properties of a blob and  a stochastically homogenous distribution of magnetic field
 advected by the same stochastic flow. This relation allows to investigate the evolution  of a localized blob at late stage when its size exceeds the viscous scale.
 It is shown that in 3-dimansional flows,  the average magnetic field of the blob increases exponentially in the inertial range of turbulence, as opposed to the late-Batchelor stage when it decreases.

Our approach reveals the mechanism of dynamo generation in the inertial range both for blobs and homogenous contributions. 
It explains the absence of dynamo in the two-dimensional case 
and its efficiency in three dimensions.
We propose the way to observe the mechanism in numerical simulations. 
\end{abstract}

\maketitle

\section{Introduction}

The advection of passive magnetic field by a turbulent flow is one of classical problems, it has been considered for more than half a century \cite{Moffat, Batchelor1950}. The problem of turbulent dynamo is needed for many physical and astrophysical applications  \cite{Parker,Rincon}. In particular, a large amount of papers is devoted to small-scale  turbulent dynamo \cite{FGV,BiferaleRev,Brandenburg12,Kulsrud92,Schekochihin,Arponen,Banerjee}. In these problems,  magnetic field fluctuations have scales much smaller than the scale at which turbulence is generated; this corresponds to inertial or viscid scale ranges of turbulence.

The most part of theoretical papers,  as well as numerical simulations, consider statistically homogenous fluctuations of magnetic field. This approximation allowed to obtain numerous results. However, the evolution of 'isolated blobs', which means localized magnetic field perturbations, in some cases differs significantly from the homogenous case \cite{ZMRS, epl18}. This 'localized' problem statement is demanded if the region occupied by  magnetic field is much smaller than the area of turbulence. But this approach may also be useful for better understanding of the origin of the dynamo mechanism.

\vspace{0.5cm}

To investigate magnetic field generation from small (seed) fluctuations, one usually considers the kinematic stage of small-scale dynamo: the magnetic field is assumed to be small and not to affect the velocity field.
There are two characteristic scales that describe a turbulent flow: the Kolmogorov viscous scale $r_{\eta}$  and the largest-eddy scale $L$. The viscous range of scales $r \lesssim r_{\eta}$  and the inertial range $r_{\eta} \ll r \ll L$ demonstrate universal behavior: the statistical properties of turbulence at these scales are independent of the geometry of the large-scale flow.

For advection problems, one assumes the velocity field to be random and to have given stationary statistical properties. Inside the viscous range, the turbulent velocity field can be considered as linear, the velocity gradient is a random function of time; this approximation is called Batchelor regime. In the inertial range, the velocity field is determined by its correlators or structure functions.

As the magnetic field is introduced in the problem, 
the presence 
of magnetic diffusivity adds the magnetic diffusion scale length $r_d$. At scales much larger than $r_d$, magnetic lines can be considered as frozen in the flow; if the width of the magnetic tube is $\sim r_d$, magnetic diffusion must be taken into account.
In the case of fully developed turbulence and large magnetic Prandtl numbers, we have
$$
r_d \ll r_{\eta} \ll L
$$
This is the range of scales we consider in this paper. We also restrict ourself to incompressible turbulence.

\vspace{0.5cm}

There are two different theoretical approaches to investigate the statistics of magnetic field advected by an isotropic homogenous turbulent flow. One of them was introduced by Kazantsev \cite{Kazantsev} and  Kraichnan \cite{Kraichnan}. It is based upon a relation between the pair correlator of magnetic field and the velocity structure function of the flow.
This approach allows to find the evolution of the two-point magnetic field correlator
for any assumption on the spatial dependence of velocity pair correlator. However, there are important restrictions: the velocity field is assumed to be Gaussian and $\delta$-correlated in time, only second order correlator of magnetic field can be found, and only statistically homogenous configurations of magnetic field can be considered.

 The other approach implies direct solution of the magnetic field evolution equation by means of the T-exponential formalism \cite{ZMRS,Chertkov, epl18}. It makes use of statistical properties of the evolution matrix and allows  to calculate the magnetic field correlators of all orders, for arbitrary (not necessarily Gaussian) velocity statistics. It also allows to consider inhomogenous, in particular, localized initial magnetic field distributions. However, this method is restricted to the Batchelor regime; it fails to produce any results in the inertial range.

The evolution of initial seed fluctuations with correlation length $l\ll r_{\eta}$ begins deep under the viscous scale, so the Batchelor regime is valid at the early stage.
For the Batchelor regime, both methods demonstrate a nice agreement: the two of them predict exponential increase
(with the same exponent) of statistically homogenous initial seed magnetic fluctuations in Gaussian $\delta$-correlated velocity field \cite{Chertkov,Kazantsev}. However, for localized initial perturbation, the 'T-exponential' method  predicts that the magnetic field would not increase: to the contrary, after some time it decreases exponentially \cite{ZMRS,epl18}. The apparent contradiction between the localized and homogeneous cases was eliminated in \cite{epl18, Scripta19} where it was shown that the evolution of homogenous initial fluctuations in Batchelor regime can be considered as a non-coherent sum of overlapping magnetic blobs.

After some time, the finiteness of the viscous scale becomes relevant, and
the inertial range velocity statistics must be taken into account.  From the Kazantsev-Kraichnan approach it follows that homogenous magnetic field continues to grow exponentially also at this later stage of evolution \cite{Kazantsev}.  This inertial-range dynamo is also observed in numerical simulations \cite{Schekochihin}.
%

But the mechanism of this increase remains obscure: at the Batchelor stage, the dynamo is a result of exponential stretching of a liquid drop; in the inertial range the exponential separation of trajectories stops (changes to a power law), so why should magnetic field grow exponentially?
On the other hand,
unlike the three-dimensional case, for the two-dimensional flow the Kazantsev-Kraichnan theory predicts exponential decay, not growth (!) at the inertial stage \cite{Kolokolov-2d}. What is the reason of the difference between two and three dimensions?
Vainshtein and Zeldovich \cite{Zeld72} proposed the idea of the stretch-twist-fold (STF) mechanism: it implies the magnetic field amplification as a result of superposition of magnetic lines during some special set of transformations. However, this idea has had no proof up to now. In \cite{Vainstein96}  the evolution of a single circular magnetic line was studied numerically, but only at initial stages the transformation similar to STF was found: after short time, the whole picture of magnetic lines  becomes too complicated to observe the STF mechanism even in a specially designed flow.

One more question is, what happens to an isolated magnetic blob as it  reaches the inertial range? No one of the two methods can answer the question.

\vspace{0.5cm}

In this paper we propose an approach that considers arbitrary stochastically homogenous magnetic field as a (non-coherent) combination of isolated magnetic blobs. This paradigm has already been used for Batchelor regime \cite{Scripta19}.
Here we apply this idea to the later stage of the evolution that corresponds to the inertial range of turbulence.  We derive the relation ('energy correspondence') between the mean-square magnetic field amplitude of the homogenous distribution and the averaged magnetic energy of one blob. This relation allows to use the results obtained for homogenous case (for instance, by means of Kazantsev-Kraichnan method) to examine the evolution of localized magnetic perturbations; in particular, to trace this evolution in the inertial range.
The result is unexpected: after exponential decay in the end of the Batchelor stage, the
exponential increase of the  magnetic field of the blob recommences at the inertial range in the case of 3 dimensions.

On the other hand, this conception gives physical interpretation of the results obtained earlier for homogenous distributions in two- and three- dimensional turbulence; it allows to analyze these distributions as a result of superposition and self-superposition of blobs.
We also suggest a physical mechanism of dynamo generation. A way to verify this conception numerically is proposed.


\vspace{0.5cm}

The paper is organized as follows. In the next Section we derive the 'energy correspondence' relation. In Section III we recall the basic idea and results of the Kazantsev-Kraichnan method. In section IV we consider the evolution equation methods; we recall the properties of the evolution matrix and its relation to the Lyapunov indeces, and derive the time evolution of the liquid drop volume and of the volume occupied by a magnetic blob in the cases of Batcherlor regime and inertial range.
In Section V we combine the results of the previous Sections to find the evolutionary properties of isolated blob. In Discussion we apply these results to explain the behavior of homogenous distribution in different cases. We show, in particular, that the difference between two and three dimensions in the homogenous case may be 
interpreted as a consequence of topological properties of an isolated blob.

In Appendix we consider Kazantsev-Kraichnan equation in details; we  present the alternative way to solve it by reduction to Fokker-Planck instead of Schr\"{o}dinger equation. This allows to find the growing solutions very easily. A particular case of this formalism is known as 'stochastic quantization'.

\section{Energy correspondence}

In the frame of kinematic regime (feedback neglected), the evolution of magnetic field advected by a turbulent flow is described by the equation:
\begin{equation}\label{1}
\frac{\dd \mathbf{B}}{\dd t} = \bigl(\mathbf{B}\nabla\bigr)\mathbf{v} -\bigl(\mathbf{v} \nabla \bigr)\mathbf{B}+  \varkappa \Delta \mathbf{B}
\end{equation}
Here $\varkappa$ is the magnetic diffusivity, ${\bf v}(t,\br)$ is the velocity of the flow. It is described by the Navier-Stokes equation, but for the purposes of advection problems $\bf v$ is normally considered as random process with given properties.
Statistics of $\bv (t,\br)$ will be considered in more details in Section IV.
 Hereafter we consider homogenous and isotropic, fully developed turbulence in incompressible fluid and high magnetic Prandtl numbers, which means that magnetic diffusivity is small as compared to viscosity $\eta$:
$$
Pr_m = \eta / \varkappa \gg 1
$$
One additional constraint comes from non-divergency condition,
\begin{equation}  \label{divB}
{\rm div} B =0
\end{equation}
We are interested in the second-order correlator,
\begin{equation}
\langle  B_i (t,\br) B_j (t,\br') \rangle = K_{ij}(t,|\br-\br'|)
\end{equation}
and, in particular, in the mean-square magnetic field $\langle B^2 (t,\br) \rangle$. The average is taken over the velocity field ensemble and over initial conditions.

 Now we concentrate on the initial (seed) magnetic field distribution.  Let $B(0,\br)$  be a random homogenous vector field with given pair correlator:
\begin{equation}  \label{BB0}
\langle  B_i (0,\br) B_j (0,\br') \rangle_{i.c.} = K^0_{ij}(|\br-\br'|)
\end{equation}
Here the average is taken over the ensemble of initial conditions. From homogeneity, isotropy and with account of (\ref{divB}) it follows that the Fourier transform of ${\bf K}^0$ satisfies
$$
K^0_{ij}(\bk) = \left( \delta_{ij} - \frac{k_i k_j}{k^2} \right) K^0 (k)
$$
and from the Bochner-Khinchine's theorem \cite{Monin-Yaglom}, $K^0(k) \ge 0$.

We present the initial magnetic field  in the form:
\begin{equation}  \label{razlojenieB0}
B_i(0,\br) = \int d\br_0 b^0_{ij}(|\br-\br_0|) C_j(\br_0) =0
\end{equation}
where $\bf C$ is a random vector function satisfying
$$
\langle C_i(\br_0) C_j(\br_0')\rangle = \frac 1{l_0^3} \delta_{ij} \delta (\br_0 - \br' _0)
$$
where $l_0$ is the correlation length of $K_{ij}^0$,
and  $b^0_{ij}$ is a non-random function determined by its Fourier transform:
$$
b^0_{ij}(\bk)= l_0^{3/2} \left( \delta_{ij} - \frac{k_i k_j}{k^2} \right) \sqrt{ K^0 (k)}
$$
One can easily check that this choice of $b^0_{ij}$ guarantees the compliance with (\ref{divB}) and (\ref{BB0}).

The equation (\ref{razlojenieB0}) can be interpreted as
as a decomposition of $\bB$ into a sum of 
independent magnetic blobs of identic shape (${\bf b}^0_{ij}$) distributed randomly in space (position of each blob is indicated by $\br_0$) with different
random weights ($|\bf C|$) and orientations (${\bf n}={\bf C}/|{\bf C}|$).

The time evolution of $\bB$ can be described by the same decomposition
\begin{equation}  \label{razlojenieB}
B_i(t,\br) = \int d\br_0 b_{ij}(t,\br,\br_0) C_j(\br_0)
\end{equation}
where $b_{ij}$ satisfies the same equation (\ref{1}) as $\bB$,
$$
\frac{\dd b_{ij}}{\dd t} = b_{mj} \frac{\dd }{\dd r_m} {v_i} - v_m \frac{\dd }{\dd r_m} b_{ij}  +\varkappa
\frac{\dd ^2}{\dd \br^2}  b_{ij}
$$
and the initial condition $b_{ij}(0, \br, \br_0)=b^0_{ij} (|\br-\br_0|)$.
We note that $b_{ij}(t,\br,\br_0)$ is a functional of ${\bf v} (t,\br)$ and  depends on both variables $\br$ and $\br_0$, not on their difference. The isotropy and homogeneity of magnetic field is destroyed by any specific realization of velocity field. This corresponds to asymmetric evolution of initially symmetric magnetic blobs. To restore the symmetry, one has to average over the velocity field.

\vspace{0.5cm}

It is easy to calculate the mean-square magnetic field of the homogenous distribution (\ref{razlojenieB}); taking the average over velocity distribution and initial conditions, we get
\begin{equation} \label{promejut}
\begin{array}{rl}
\langle B^2 \rangle _{i.c.,{\bf v}} =& \left \langle \int d\br_0 b_{im}(t,\br,\br_0) C_m(\br_0) \int d\br'_0 b_{ik}(t,\br,\br'_0) C_k(\br'_0) \right \rangle_{i.c.,\,\bv} \\
=& \displaystyle{ \frac 1{l_0^3} \left \langle \int d\br_0 Tr {\bf b}^T {\bf b}(t,\br,\br_0) \right  \rangle _{\bv} }
\end{array}
\end{equation}
For each $\br_0$, $\langle {\bf b}^T {\bf b} \rangle_{\bv} $  describes the evolution of one particular blob of a given initial amplitude.
From homogeneity of the velocity field it follows that the average depends only on $\br - \br_0$:
%
\begin{equation} \label{btb-promejut}
\langle {\bf b}^T {\bf b} \rangle_{\bv} = \langle {\bf b}^T {\bf b} \rangle_{\bv} (t, \br-\br_0)
\end{equation}
Now, in (\ref{promejut}) we replace the integral over $\br_0$ with the integration over $\br$.
Thus, we obtain
\begin{equation} \label{B2general}
\langle B^2 \rangle _{{\bf v,}\, i.c.} = \frac 1{l_0^3} \int d\br Tr \langle {\bf b}^T {\bf b}\rangle _{\bv}
\end{equation}

\vspace{0.5cm}

One and the same blob undergoes the same evolution independently of its environment, both if it is alone or a part of some set of blobs. So, we now consider  an isolated magnetic blob with magnetic field $b_{ij}(t,\br,\br_0) n_j$,  here $\bf n$ is some realization of  randomly oriented unit vector.  The energy of the blob is
$$
E_{blob}= \int d\br \left( b_{ij}(t,\br,\br_0) n_j \right) \left( b_{im}(t,\br,\br_0) n_m \right)
$$
In this expression, there is no averaging. Now we average over the velocity ensemble:
$$
\langle E_{blob} \rangle _{\bv} = \left( \int d\br \langle {\bf b}^T {\bf b}(t,\br,\br_0) \rangle_{\bf v} \right)_{jm} n_j  n_m
$$
From (\ref{btb-promejut}) it follows that the integrand depends only on the difference $\br-\br_0$, so the integral does not depend on $\br_0$. Hence,
the  tensor inside the round brackets has no preferential directions and, thus, is proportional to $\delta_{jm}$. Eventually, we get
\begin{equation}  \label{Eblob}
\langle E_{blob} \rangle _{\bv} (t) = \frac 13  \int d\br Tr \langle {\bf b}^T {\bf b} \rangle_{\bf v}
\end{equation}

Comparing (\ref{Eblob}) to (\ref{B2general}), we obtain:
\begin{equation} \label{EC}
\langle E_{blob}   \rangle _{\bv}  = \frac 13 \langle B^2 \rangle _{i.c.,{\bf v}}  l_0^3
\end{equation}

In what follows we refer to this equation as to energy correspondence. It is valid for all regimes independently of the velocity statistics. This equality establishes the relation between the problems of evolution of separate blob and of homogenous distribution of fluctuations. According to (\ref{razlojenieB}), for any homogenous initial distribution one can choose a blob of corresponding shape so that they evolutionate concordantly, and (\ref{EC}) is valid.  The shape of the blob is determined by the correlation function of initial magnetic field (\ref{BB0}).
And vice versa, for any individual magnetic blob there exists the 'corresponding' homogenous distribution such that (\ref{EC}) holds.

This allows to apply the results obtained for homogenous configurations to analyze the evolution of a separate blob.

\section{Homogenous fluctuations: Kazantsev-Kraichnan  theory}
Many numerical and analytical investigations demonstrate growth of homogenous fluctuations at large magnetic Prandtl numbers \cite{Rincon, Schekochihin}. The first and the most popular theory was proposed by Kazantsev \cite{Kazantsev}.

The energy correspondence is valid for arbitrary homogenous and isotropic velocity statistics. However, in order to apply the Kazantsev-Kraichnan theory, we have to restrict ourself to Gaussian statistics and $\delta$-correlated in time second-order velocity correlator:
$$
\left \langle v_i(t,\br) v_j(t,\br') \right \rangle = D_{ij}(\br-\br') \delta(t-t')
$$
The condition of isotropy and non-divergency requires that $D_{ij}$ takes the form
$$
D_{ij} ({\bf x})= 2P(x) \delta_{ij}+ x P'(x) \left( \delta_{ij}-\frac{x_i x_j}{x^2} \right)
$$
Analogously, from homogeneity, isotropy and non-divergency of the magnetic field it also follows
$$
\begin{array}{ll}
\langle B_i(t,\br) B_j(t,\br+{\bf x}) \rangle &= K_{ij} (t, x)  \\
&=2G(t,x) \delta_{ij}+ x G'_x(t,x) \left( \delta_{ij}-\frac{x_i x_j}{x^2} \right)
\end{array}
$$
The Kazantsev-Kraichnan equation establishes the functional relation between these correlators:
\begin{equation} \label{KK}
\frac{\partial G}{\partial t} = 2(\varkappa + \frac 14 S_2)G'' + \left( 8\frac{\varkappa+\frac 14 S_2}r
+ \frac 12  S_2' \right) G' +
 \left( \frac 12 S_2''+\frac 2r S_2' \right) G \ ,
\end{equation}
where
$$
S_2(r) = \left \langle \left( v_{\parallel}(0)- v_{\parallel}(r) \right)^2 \right \rangle
= 4 \left( P(0)-P(r) \right)
$$
is the longitudinal velocity structure function.
The viscous and inertial  ranges of scales correspond to different asymptotes for $S_2$:
\begin{equation} \label{S2-asympt}
S_2(r) \propto \left\{ \begin{array}{l}  r^2 \ , \ \ r\ll r_{\eta} \\
                                        r^{2-\zeta} \ , \ \ r_{\eta} \ll r \ll L \\
                                        r^0 \ , \ \ r \gg L
\end{array} \right.
\end{equation}

In the Batchelor regime, the approximation $S_2\propto r^2$ is valid for all $r$; the equation (\ref{KK}) has the growing mode\cite{Kazantsev} (see Appendix A for more details):
\begin{equation} \label{5/2D}
G \sim e^{\gamma_B t} \ , \ \ \gamma_B=\frac 52 D
\end{equation}
Here $D=Tr D_{ij}(0) = 3 P(0)$.

  So, the mean-square magnetic field increases exponentially in the Batchelor regime. According to (\ref{EC}), the average energy of one single blob must grow with the same increment. This result coincides with the solution for the blob evolution obtained in \cite{ZMRS, epl18}.

  In general case, the evolution of magnetic field is determined by the whole expression (\ref{S2-asympt}).
It appears \cite{Vincenzi} (see also Appendix A) that in the three-dimensional turbulence the Kazantsev-Kraichnan equation still has the growing mode close to that of the Batchelor case:
\begin{equation} \label{gamma-inertial}
\gamma = \gamma_B + O(r_d/r_{\eta})
\end{equation}
However, this is not an inevitable rule: in the case of  two dimensions, the two-dimensional analog to Eq.(\ref{KK})  has the growing mode in the Batchelor regime but
has no  
growing solutions
in the case of finite $r_{\eta}$. So, in two-dimensional flows, the exponential increase of mean-square magnetic field at early ('Batchelor') stage of evolution is followed by  a power-law decay in the long-time asymptotics \cite{Kolokolov-2d}. According to (\ref{EC}), so does the energy of a single two-dimensional blob.
 For a blob, this transition happens when the  size of the blob becomes larger than $r_{\eta}$. When does it happen in the homogenous distribution? The answer follows from (\ref{EC}): the exponential growth stops when the 'corresponding' (in terms of (\ref{razlojenieB})) blobs reach the inertial range.

\section{ Volume evolution: T-exponentials }

According to the energy correspondence, the average magnetic energy of an isolated three-dimensional magnetic blob would  increase with the increment (\ref{gamma-inertial}), even though the size of the blob becomes larger than the viscous scale. Does the magnetic field inside  the blob increase or decay? To answer this question, one can present the energy of the blob in the form
$$
\left \langle E_{blob} \right \rangle \sim \left \langle b^2  V \right \rangle
$$
where $V(t)$ is the  volume occupied by the blob. So, to understand what happens to the magnetic field, we have to analyze the evolution of the blob volume.   

\subsection{Liquid drop: Batchelor regime}
For the beginning, we consider the evolution of some liquid drop (not magnetic field). In the viscous range of scales, the stochastic  velocity field can be approximated by a linear function and is described by its random velocity gradient tensor $A_{ij}(t)=\d v_j (\bR,t)/ \d r_i$ where $\bR(t)$ is the position of center of the drop. The statistics of this tensor is assumed to be known. The separation of trajectories of two fluid particles
is described by the equation
\begin{equation} \label{drdt=Ar}
 d {\bf \delta r} / dt = {\bf A} {\bf  \delta r}
\end{equation}
The formal solution of this equation can be written by means of the T-exponent:
$$
{\bf \delta r }(t) = {\bf Q} {\bf \delta r }(0) \ , \ \ {\bf Q}=T exp \int \limits_0^t {\bf A}(t) dt
$$
The long-time behavior of the evolution matrix $Q$ has been investigated in many papers (see, e.g., \cite{Oseledets,Let-survey,CrisantiPaladinVulp,JOSS1}).
The multiplicative ergodic theorem \cite{Oseledets}
 states that with unitary probability there exists the limit
\begin{equation}  \label{lambda}
\lambda_i = \lim \limits_{T \to \infty} \frac{\ln d_i(T)}{T} \ , \qquad \lambda_1 \ge \lambda_2  \ge \lambda_3
\end{equation}
where $d_1,d_2,d_3$ describe the evolution of an arbitrary initial basis ${\bf e_1},{\bf e_2},{\bf e_3}$:
\begin{equation}  \label{d123}
\begin{array}{l} \displaystyle
d_1(t) = |{\bf Q}(t) {\bf e}_1| \ , \ \
d_2(t) = \frac{\left| \left[ {\bf Q}(t) {\bf e_1} , {\bf Q}(t) {\bf e_2} \right] \right|}
{|{\bf Q}(t) {\bf e_1}|} \ , \ \ \\
d_3(t) = \frac{\left| \left( {\bf Q}(t) {\bf e_1} ,{\bf Q}(t) {\bf e_2} , {\bf Q}(t) {\bf e_3}\right) \right|}
{\left| \left[ {\bf Q}(t)  {\bf e_1} ,{\bf Q}(t) {\bf e_2} \right] \right|}
\end{array}
\end{equation}
The set of constants $\lambda_i$ is called Lyapunov spectrum \cite{Oseledets}. It is an important statistical characteristic of the process $A(t)$, and it does not depend on the realization.
From incompressibility of the fluid it follows that ${\det}{\bf Q}=1$ and  hence
$$
 d_1 d_2 d_3 = 1 , \ \ \lambda_1+\lambda_2+\lambda_3=0
$$


If the initial drop was an ellipsoid, it remains ellipsoidal during the whole Batchelor period of evolution; but the ratio of its axes changes, as well as their orientation.
Because of  (\ref{lambda}), after long enough time there are three possibilities depending on
the sign of $\lambda_2$:
\\
if $\lambda_2<0$ there is one exponentially stretching axis and two contracting ones, the drop looks like a stretching and rotating  'stick' ('filament' is the conventional term for this situation, but here we stress that the construction is  inflexible); the length of the 'stick' is proportional to $d_1$;
\\
if $\lambda_2>0$ there are two exponentially stretching directions, we get a 'pancake' or, rather, a 'sheet pan';
according to \cite{lambda2=1/4}, this is the case of the Navier-Stokes turbulence.
\\
if $\lambda_2=0$ there is one stretching and one contracting directions, the third axis is stabilized and does not change significantly ('spaghetti');  the case corresponds to time-invariant flows, in particular, this is the case of the Gaussian velocity correlator in Kazantsev-Kraichnan model.

From (\ref{d123}) it follows that in all these cases,  the length of arbitrary segment advected by a flow is (with unit probability) proportional to $d_1$; the square of an arbitrary infinitesimal fragment of surface changes as $d_1 d_2$:
\begin{equation} \label{length and square}
 {\delta \mathcal L}     \propto d_1 (t) \ , \ \ {\delta \mathcal S} \propto d_1 d_2 (t) = d_3 ^{-1} (t)
\end{equation}
 Any advected volume is constant, as it should be in incompressible fluid, since $d_1 d_2 d_3 = 1$.

\subsection{Liquid drop: Inertial range}
The linear approximation for velocity field stops working after
\begin{equation} \label{vremya-visc}
t \sim \frac 1{\lambda_1} \ln \frac {r_\eta}{l_0}
\end{equation}
 where $l_0<r_{\eta}$ is the initial size of the drop: after this time, the longest axis of the ellipsoid exceeds the viscous scale.  After this time, viscosity is still essential at distances smaller than $r_{\eta}$, so, in the vicinity of any local point  the velocity field remains linear, and exponential stretching and contraction continues; however, in different regions separated by distances larger than $r_{\eta}$, the  stretching occurs in different  directions; so, 'stick' transforms to a bending 'rope' (or a filament), and 'sheet pan' becomes a flexible  'towel'.

Consider the evolution  of  a line (filament)  advected by the flow. Let its length belong to the inertial range, i.e., ${\mathcal L} \gg r_{\eta}$.
Eq.(\ref{length and square}) remains still valid for each small part of the curve, and its integral length is not random and is  proportional to the average over $d_1$:
\begin{equation} \label{dlina-inertial}
{\mathcal L}  \simeq \langle d_1 \rangle_{\bv}(t) {\mathcal L_0}
\end{equation}
 To prove this,
  denote the position of any point of the line by $\br (\alpha,t)$ where $\alpha$ is the parameter along the line. Then
$$
\d \br (\alpha,t) / \d t = {\bf v} (\br(\alpha,t),t)
$$
where the velocity field is assumed to be known.
The length of the line is
$$
{\mathcal L}  =   \int d\alpha \left| X(\alpha,t) \right|
$$
where $ {\bf X}(\alpha,t)= \d \br (\alpha,t) / \d \alpha $; the evolution equation for $\bf X$ is
$$
\frac{\d {\bf X}}{\d t} = \frac {\d {\bf v}}{\d \alpha} = \frac{\d {\bf v}}{\d \br} {\bf X} =  {\bf A(\br(\alpha,t),t)} {\bf X}
$$
This equation coincides with (\ref{drdt=Ar}), so $\bf X$ undergoes the same evolution as an infinitesimal fluid segment;  the solution takes the form
$$
{\bf X}(\alpha,t)= {\bf Q} (\alpha,t) {\bf X}(\alpha,0) \ , \ \ {\bf Q}=T \exp \int \limits_0^t dt {\bf A} (\br(\alpha,t),t)
$$
Thus, just as in (\ref{length and square}), we get
$ |{\bf X}| = d_1 (t, \alpha)$ where the realization of $d_1$ depends on the trajectory.
Statistics of $\bf A$ (and hence of $d_i$) is the same along all the trajectories because of homogeneity of the flow.  Now, since the length of the line is much larger than the viscous scale, its different parts evolve independently, and from central limit theorem it follows that the length is not random; it is proportional to the ensemble average of $|\bf X|$,
$$
 {\mathcal L}  = \int d \alpha d_1(t,\alpha) |{\bf X}(\alpha,0)| \simeq \langle d_1(t) \rangle  {\mathcal L}(0)
$$
and we arrive at (\ref{dlina-inertial}).

\vspace{0.3cm}
%
%
%
%

In a similar way one can check that the area of any small surface element advected by the flow changes as $d_1 d_2 = 1/d_3$, just as in the Batchelor regime; and the integral square of a large (compared to $r_{\eta}$) surface is
\begin{equation} \label{ploschad-inertial}
{\mathcal S}  \simeq \langle d_3 ^{-1} \rangle_{\bv}(t) {\mathcal S_0}
\end{equation}

So, the length of a line and the square of a surface continue to increase exponentially even after they  become larger than the viscous range. However, the exponential separation of particles' trajectories stops as soon as the line's length gets into the inertial range. Thus, different points of the line do not move away one from another faster than a power-law function of time, and the whole line remains inside some sphere, the radius of the sphere not increasing exponentially. Hence, after some time the filament would 'fill' the sphere very 'tightly'. All the same is true for a surface.

 \subsection{Magnetic blob: Batchelor regime}

Now we proceed to 
a localized perturbation of magnetic field. Let the initial scale of the perturbation lie in the limits
$$
r_d < l_0 \ll r_{\eta}
$$
At the early stage of evolution, the magnetic blob is entirely frozen in the accompanying flow.
This regime is called 'ideal conductor', and evolution of magnetic blob's volume coincides with the evolution of a liquid drop during this stage.  If the initial blob was ellipsoidal, it remains ellipsoidal until the end of the
Batchelor regime, although the shape of the ellipsoid and its orientation change as time goes.

This ideal conductor stage continues until the smallest of the blob's diameters becomes of the order of $r_d$. This happens at
\begin{equation} \label{t-id-cond} 
t \sim |\lambda_3|^{-1} \ln \left( l_0/ r_d \right)
\end{equation} 
 After that, the magnetic diffusion prevents the blob from further contraction. At scales larger than $r_d$, the magnetic field remains frozen in the flow, and the exponential stretching along the directions corresponding to $d_i>1$ continues, but the contraction along the other directions stops. So, magnetic blobs look similar to fluid drops, but the 'magnetic' filaments or pancakes  have finite thickness $\sim r_d$.

With account of (\ref{length and square}), one derives that the volume occupied by the magnetic field is the product of its length/square and its cross section/thickness:
$$
 V \sim r_d^2 {\mathcal L} \sim \frac{r_d ^2}{l_0 ^2} d_1 V_0
$$
for filaments and
$$
V \sim r_d  {\mathcal S} \sim \frac{r_d}{l_0 } d_3^{-1} V_0
$$
for pancakes.
The T-exponential  approach \cite{Chertkov, epl18} allows to find the magnetic field in the middle of the blob:
$$
 b^2 (t) \propto \exp(-2 |\ln d_2|) \ ,
$$
the expression is valid both for filaments and pancakes. The magnetic field decreases in Batchelor regime after (\ref{t-id-cond}); but the energy of a blob increases \cite{ZMRS}. Multiplying $b^2$ by the volume and taking the average over velocity field, one can find the time dependence of the average blob energy and compare the result to the mean-square magnetic field in the homogenous distribution. This was done in \cite{Scripta19}, and
not only (\ref{EC}) is confirmed but also all statistical moments of these two quantities coincide (up to the pre-exponent) in Batchelor regime. In particular,
\begin{equation}   \label{B2 homogen-log}
 \frac 1t \langle \ln E_{blob} \rangle_{\bv} = \frac 1t \langle \ln \langle B^2 (t) \rangle_{i.c.}\rangle_{\bv} = \left\{ \begin{array}{ll} \lambda_1-\lambda_2 & \ \mbox{if} \  \lambda_2>0 \ , \\
\lambda_2-\lambda_3 & \ \mbox{if} \  \lambda_2<0  \end{array}  \right.
\end{equation}
Also, both $\langle E_{blob} \rangle$ and $B^2$ have the exponents equal to (\ref{5/2D}), so all the results obtained by different methods are concordant.

 \section{Inertial stage of evolution of three-dimensional magnetic blob}

As the length of the blob exceeds the viscous scale, it continues to follow the accompanying fluid drop.
 Every small part of the blob stretches exponentially in the local linear velocity field, although the direction of stretching and the extension rate  depend on the region. The only difference from a fluid drop is that the magnetic blob cannot become thinner than $r_d$: it remains the same thickness, while the fluid drop continues its contraction.
Thus, the transversal sizes of either filament or pancake remain equal to $r_d$, and their longitude or area increase exponentially in accordance with (\ref{dlina-inertial}),(\ref{ploschad-inertial}).

So, it seems that nothing is changed as the magnetic blob passes from viscous to the inertial range of scales: its volume continues to increase exponentially as well as its average energy.
This corresponds to the behavior of $\langle B^2 \rangle$ in the homogenous case; both theory (\ref{gamma-inertial}) and simulations \cite{Schekochihin} predict the continued exponential increase for three dimensional flows.

However, we recall that the exponential separation of any two points of the blob stops at the inertial range.
So, just as a fluid drop, the magnetic blob remains inside some sphere, the radius of the sphere increasing slowly (power law). But, as opposed to a fluid drop, the volume of magnetic blob does not remain constant:
 the length (area) increases exponentially while the cross section (thickness) remains constant as a result of diffusion. So, after some time the whole sphere would be filled up by the bent and folded 'blob'.  This time  can be easily estimated assuming the sphere's diameter $\sim r_{\eta}$; it is
$t \simeq 2 {\lambda_1}^{-1} \ln (r_{\eta}/r_d)$ for a 'filament' and
$t \simeq  {|\lambda_3|}^{-1} \ln (r_{\eta}/r_d)$ for a 'pancake'.  In both cases it is comparable to the duration of the viscous stage (\ref{vremya-visc}).

What happens after this time? The local stretching continues but there is no more room inside the sphere; so, different parts of the same blob start to overlap. The exponential increase of volume stops from this moment.

But the stochastically homogenous magnetic field continues to grow; because of (\ref{EC}), so does the average blob energy. Hence, we arrive at the conclusion that the mean-square magnetic field of a three-dimensional isolated blob must increase exponentially at this stage. In particular, for Gaussian $\delta$-correlated flow, with account of (\ref{5/2D}) we get
$$
\left \langle b^2 \right \rangle \propto \exp \left({\frac 52 D t }\right)
$$

\section{Discussion}

So, the evolution of an isolated magnetic blob can be divided into three parts: the 'ideal conductor' stage
($ t < t_{id.cond} = |\lambda_3|^{-1} \ln (l_0/r_d) $)
 when the magnetic lines are frozen in the fluid flow and diffusion scale is negligible as compared to the blob's tranverse size; viscous, or Batchelor, stage ($ t < t_{visc} = \lambda_1^{-1} \ln (r_{\eta}/l_0) $) when the length (largest size) of the blob is smaller than the Kolmogorov scale, and the velocity field can be considered as linear; and the inertial stage  when the integral length of the blob exceeds $r_\eta$, although the whole folded blob remains inside an area with the size comparable to $r_{\eta}$.
Now, because of the decomposition (\ref{razlojenieB}), the same division into three stages appears to be applicable to an isotropic magnetic field distribution, $l_0$ being the initial correlation length instead of the initial blob scale.

Here we restrict our consideration to the condition that the ideal conductor stage ends before the inertial stage starts, $t_{id.cond} \ll  t_{visc}$. This means that the transverse size of the blob becomes comparable to $r_d$, and after that its length becomes comparable to $r_{\eta}$.  This condition corresponds to
$l_0^{\lambda_1+|\lambda_3|}  \ll r_{\eta}^{|\lambda_3|} r_d^{\lambda_1} $ (or, roughly, $l_0 \ll \sqrt{ r_{\eta}r_d}$).

There are two important results obtained in the previous Section. First, we discover that the magnetic field of a localized magnetic perturbation (blob) increases exponentially at the inertial stage of its evolution,
 after the time  $ t \sim max(\lambda_1,|\lambda_3|)^{-1} \ln (r_{\eta}/r_d) $.
This is new and unexpected   because
the magnetic field of a blob is known to decrease during the  earlier (viscous) stage.

Second, an important result is that  at the inertial stage of evolution, different parts of the same blob necessarily overlap, and the effective volume of this overlappings increases exponentially.  Simple as it is, this fact provides a key to the mechanism responsible  for the growth of magnetic field.  It reminds  the stretch-twist-fold (STF) mechanism proposed in \cite{Zeld72}. But the literal understanding of the STF procedure implies the presence of some special kind of transformations that would realize the 'twist' and 'fold'. To the contrary,
here we see that 'twist' is not necessary, while 'fold' occurs automatically.  As two parts of a blob overlap, the magnetic field increases if the angle between the two magnetic inductions is smaller than $\pi/2$. As a result of interference, $b^2$ increases on average in each act of overlapping. This is enough to produce the exponential increase.

This  mechanism not only explains the increase of magnetic field in an isolated blob; by means of (\ref{razlojenieB}) it  also reveals  the cause of the dynamo effect in isotropic turbulence.
At early (viscous) stage, as it was shown in \cite{Scripta19}, the increase of statistically homogenous fluctuations is a result of incoherent summation of magnetic fields produced by different blobs
at every point.
To the contrary, at the inertial range the exponential increase of statistically homogenous field is produced by coherent overlapping of different parts of the same blob (Table~\ref{Table1}).

We stress that, unlike the viscous stage of evolution, the increase of total magnetic field is caused by magnetic field growth of each blob, not just by their summation. So, the mechanism responsible for the dynamo is quite different during the viscous and the inertial stages.

\vspace{0.2cm}
%
%
\begin{table}[b]
\caption{Evolution  of magnetic field and the mechanism of dynamo generation at different stages. The new results obtained in this paper are set off in boldface. \newline
}
\label{Table1}%
\begin{ruledtabular}
\begin{tabular}{|c|c|c|c|}
\hline
 stage      &  blob       & homogenous & mechanism   \\
            &             &  distribution     & of dynamo   \\
\hline
viscous    & $V=const$      &   $\langle B^2 \rangle \uparrow $ exp & compression \\
id. cond.   &$\langle b^2 \rangle \uparrow $ exp &  & of magnetic tubes          \\
\hline
viscous      &  $V \uparrow$ exp    &   $\langle B^2 \rangle \uparrow $ exp & uncorrelated  \\
2d,3d        &  $\langle b^2 \rangle \downarrow $ exp &   & overlap of blobs \\
\hline
inertial      &  $V \uparrow$ power    &$\langle B^2 \rangle \uparrow $ exp & { \bf{correlated} } \\
3d           & $ \boldsymbol{ \langle b^2 \rangle \uparrow} $ {\bf exp}&   & \bf self-overlap \\
\hline
inertial      &  $V \uparrow$ power    &$\langle B^2 \rangle \downarrow$ power & \bf{ anticorrelated}  \\
2d           &  $\boldsymbol{ \langle b^2 \rangle \downarrow }$ {\bf power} &   & \bf self-overlap \\
\hline
\end{tabular}
\end{ruledtabular}
\end{table}
\vspace{0.2cm}

Consider now isotropic magnetic field in two-dimensional turbulence. Unlike the three-dimensional case, 2d turbulent dynamo ceases at the inertial stage: the Kazantsev equation has no growing mode, and the increase of magnetic field at viscous stage of evolution is followed by power-law decay. 
From (\ref{EC}) we conclude that in this case, the mean-square  field of each blob decays, too. How does this agree with the conception of self-overlapping blobs?

The answer lies in the topological structure of magnetic lines in
two and three dimensions.
In two-dimensional case, strong bending of magnetic line  necessarily means that oppositely directed segments of the line approach each other (Fig.1a). As two parts of a blob overlap, their reconnection terminates the magnetic field. Similar idea was suggested in~\cite{Kolokolov-2d}. So, in two dimensions the self-overlapping is always anti-correlated for topological reasons, while in the 3d case
codirectional magnetic lines configurations are as probable as contradirectional ones (Fig.1b), and correlated self-overlapping is efficient.
\\
\begin{table*}
\caption{Stages of evolution  of a 3-dimensional magnetic blob. \newline
}
\label{Table2}
\begin{ruledtabular}
\begin{tabular}{|c|c|c|c|}
\hline
 time      &  what happens       & blob volume & magnetic field  \\
\hline
\hline
$t \lesssim t_{id.cond}=\frac 1{|\lambda_3|} ln \frac {l_0}{r_d}$    &  magnetic lines frozen in the flow;     &   const & exp increase\\
                                                            & contraction,  exponential stretching & & \\
\hline
$t_{id.cond} \lesssim t \lesssim t_{visc}=\frac 1{\lambda_1} ln \frac {r_{\eta}}{l_0}$  & constant transverse size, & exp increase & exp decrease           \\
                                                            &   exponential stretching & & \\
\hline 
$t_{visc}\lesssim t \lesssim t_{fill} = \frac 1{max(\lambda_1,|\lambda_3|)} \ln \frac {r_{\eta}}{r_d}$  & the blob bends, folds, & exp increase &   exp decrease        \\
                                                            &   fills a sphere $\sim r_{\eta}$ & & \\
\hline 
$ t \gtrsim t_{fill}$  & self-overlaps of the blob  & power increase &   exp increase        \\
\end{tabular}
\end{ruledtabular}
\end{table*}

\begin{figure}[h]
\hspace*{-0.5cm}
\includegraphics[width=9cm]{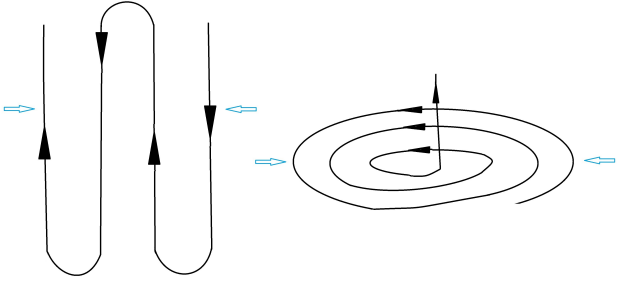}  
\caption{Illustration to anti-correlated (a) and correlated (b) overlap of magnetic lines.}
\end{figure}

Finally, the proposed mechanism of magnetic field growth can be observed in numerical simulations. It has always been difficult to trace the evolution of a magnetic tube even in 'artificial' turbulence. Our proposition is to consider an isolated magnetic blob with initial size $r_d \ll l_0 \ll \sqrt{r_{\eta} r_d}$ advected in a turbulent flow.
According to the results obtained above, the magnetic field of the blob would undergo exponential increase at the 'ideal conductor' stage, then exponential decrease at the 'viscous' stage (Batchelor regime), and then exponential growth again.
More accurate list of different stages is presented in Table~\ref{Table2}.
Also, it may be easier to observe the self-overlapping of the blob and to confirm the increase of magnetic field as a result of the reconnection.

\begin{acknowledgments}
The authors are grateful to Prof. A.V. Gurevich for his permanent attention to their work.
The work of AVK was supported by the RSF  grant 20-12-00047.
\end{acknowledgments}

\appendix
\section{Existence of growing solutions of parabolic equations}

Consider general linear parabolic partial differential equation:
\begin{equation} \label{A1}
\frac{\d}{\d t} G =- u(\br) G - \frac{\d}{\d r} \left( \lambda(r) G \right)
+ \frac 12 \frac{\d^2}{\d r^2} \left( \sigma^2(r) G \right)
\end{equation}
$$  r>0 \ , t \ge 0 $$
Here $u(r)$, $\lambda(r)$, $\sigma^2(r)$ are given functions. We are interested in the asymptotic  behavior of solutions as $t\to \infty$.

A conventional way to solve this equation is to reduce it to the Schr\"{o}dinger equation with imaginary time  and time-dependent mass. Then one searches (analytically or numerically) for the lowest energy level; it indicates whether the growing solution of (\ref{A1}) exists or not. Here we propose another method based on the reduction (\ref{A1}) to the Fokker-Planck equation,
\begin{equation} \label{A4}
\frac{\d}{\d t} \rho(r,t)  =  \frac{\d}{\d r} \left( \mu(r) \rho(r,t) \right)
+ \frac 12 \frac{\d^2}{\d r^2} \left( \sigma^2(r) \rho (r,t)\right)
\end{equation}
For Schr\"{o}dinger equation, which corresponds $\lambda =0$,$\sigma^2=1$, this reduction to the Fokker-Planck equation is known as stochastic quantization.

The equivalence of the equations (\ref{A1}) and (\ref{A4}) can be established by means of
\begin{equation} \label{A5}
\frac{d}{dr} \left( \frac{\mu(r)}{\sigma^2(r)} \right) + \left( \frac{\mu(r)}{\sigma^2(r)} \right)^2 =
\frac{2u(r)}{\sigma^2(r)} + \frac{d}{dr} \left( \frac{\lambda(r)}{\sigma^2(r)} \right) +
\left( \frac{\lambda(r)}{\sigma^2(r)} \right)^2
\end{equation}
Then, the  solutions of these two equations are related by
\begin{equation} \label{A6}
G(r,t)= \rho(r,t) \exp \left[ \int \frac{\lambda(r)-\mu(r)}{\sigma^2(r)} dr \right]
\end{equation}
But the Fokker-Planck equation has no growing modes: actually, its Green function has stationary asymptote
$$ U(r,t|r_0,t_0) \xrightarrow[{t \to \infty}]{} \frac{\rho_0}{\sigma^2(r)} \exp \left[ 2 \int 
\frac{\mu(r)}{\sigma^2(r)} dr \right]
$$
and obeys the normalization condition
$ \int U(r,t|r_0,t_0) dr = 1$.

So, if there exists a solution $\mu(r)$ of  Eq. (\ref{A5})  for all $r>0$, then  for any $G(r,t)$ there exists the corresponding  $\rho(r,t)$, and hence, no growing solutions of Eq.(\ref{A1}) can exist. And vice versa, if Eq.(\ref{A1}) has growing modes then Eq.(\ref{A5}) has no real solutions existing on the whole $r>0$ semi-axis.

One can present this condition in more convenient form:  consider the function
$$
y(r)=\exp \left( \int \frac{\mu(r)}{\sigma^2(r)}dr \right)
$$
Then (\ref{A5}) takes the form
\begin{equation} \label{A10}
y''(r) = 2 u_{eff} (r) y(r) \ ,
\end{equation}
where $u_{eff}$ is defined by
\begin{equation} \label{A11}
2u_{eff}(r) = \frac{2u(r)}{\sigma^2(r)} + \frac d{dr} \left(  \frac{\lambda(r)}{\sigma^2(r)} \right) +
 \left(  \frac{\lambda(r)}{\sigma^2(r)} \right)^2
\end{equation}
The existence of real solution of Eq.(\ref{A5}) is equivalent to the existence of positive solution of Eq.(\ref{A10}).
One can see that \\
{\it A growing solution of (\ref{A1}) exists if and only if Eq.(\ref{A10}) has no real positive solutions. }

With this criterion one can not only determine if there exist growing modes or no, but also find the maximal  increment of growth or the minimal damping decrement.

\vspace{0.7cm}
{\bf Damping decrement}

Suppose that there are no growing modes, and the slowest-damping mode is $G(t,r) \sim \exp (-E_0 t)$, $E_0>0$.
Then the corresponding equation for $G(r,t)\exp (E_0 t)$ must have no exponentially decreasing solutions, its damping decrement is zero. So, to find $E_0$, we add the term $+\delta G$ in the right-hand side of (\ref{A1}); this corresponds to the change of variable $G(r,t)$ to $G(r,t) \exp (\delta t)$.
This added term results in the
substitution $u(r)-\delta $ for $u(r)$ in (\ref{A5}) and (\ref{A11}).
We then increase the parameter $\delta$ beginning with 0 until the positive solutions of the modified (\ref{A10}) disappear at some $\delta_0>0$.  This means that the modified equation   has a growing mode for any $\delta>\delta_0$, and has no growing solutions for $\delta<\delta_0$. So, the damping decrement of (\ref{A1}) is
$E_0 = \delta_0 $.

\vspace{0.7cm}
{\bf  Growth increment}

Analogously, let the fastest growing mode be $G\sim \exp \left( -E_0 t \right)$, $E_0<0$.
Then we add the term $-\gamma G$ into the right-hand side of (\ref{A10}), with corresponding change $u \rightarrow u+\gamma$ in (\ref{A5}) and (\ref{A11}), and we increase $\gamma$ from zero until the positive solutions disappear. The corresponding $\gamma_0 = -E_0$ is the increment of growth of (\ref{A1}).

\vspace{0.5cm}

Now we  apply this method to consider the behavior of passive scalar and vector fields advected by a turbulent flow.

\subsection{ Passive scalar}
Let $Q(r,t) = \langle n(r,t) n(0,t) \rangle $ be the pair correlation function of the scalar density $n$.
Its evolution in a turbulent flow is described by the Kraichnan equation:
$$  
\frac{\d}{\d t} Q =
\left(  \frac{4S(r)}r +2S'(r) \right) Q' + 2S(r) Q''
$$    
Here $S(r) = \varkappa + \frac 14 S_2(r)$, and $S_2$ is the longitudinal velocity structure function of the flow.
Scalar density decreases as a function of time; to calculate the damping decrement, we add the term $\delta Q$
in the equation. Then the effective potential in (\ref{A10}) is equal to
$$
2 u_{eff}^{sc}= -\frac{\delta}{2S(r)} + \frac 12 \frac{d^2}{dr^2} ln S(r) + \frac 1r \frac{d}{dr} ln S(r) +
\frac 14 \left( \frac d{dr} ln S(r) \right) ^2
$$
According to (\ref{S2-asympt}), there are different asymptotes for $S_2$ in different scale ranges.

For the {\bf Batchelor  regime} we have
$$
S_2 = \frac 43 D r^2
$$
Then, as $r \gg \sqrt{\varkappa/D}$,  the effective potential can be approximated by
$$
2u_{eff}^{sc} \simeq \left( 2 - \frac 3{2D} \delta \right) /r^2
$$
The asymptotic solution of (\ref{A10}) in the case is
\begin{equation} \label{y}
y = A r^{a_1} + B r^{a_2}
\end{equation}
where $a_1,a_2$ are the roots of the equation $a(a-1)=2-(3/2D)\delta$:
$$
a_{1,2} = \frac 12 \pm \sqrt{\frac 2{3D} \left(\frac{3D}2 -\delta \right)}
$$
We see that for $\delta>3D/2$ the solution (\ref{y}) oscillates, so there are no positive solutions of (\ref{A10}).
Thus, $\delta_B = 3D/2$ is the exact damping decrement of the passive scalar. This result coincides with that obtained in \cite{BF,Kraichnan}.
One can check that for any $\delta \le 3D/2$  the positive solution exists, so there are no solutions decreasing slower than $\delta_B$.

In the {\bf inertial range} we have for $S_2$, and hence, for $S$ the asymptote
$$
S(r) \sim r^{2-\epsilon}  \ , \ \  r \gg r_{\eta}
$$
The effective potential is $2u_{eff}^{sc} \sim -\delta / r^{2-\epsilon}$.
For any $\delta$, for $\epsilon >0$ the solution of (\ref{A10}) oscillates, i.e., there are no positive $y$.
Thus, we arrive to the conclusion that maximal damping decrement in the inertial range is $\delta_{(in)} =0$, and the decrease of scalar field density correlator at the inertial stage of evolution obeys a power law \cite{Kraichnan}.  In accordance with the scalar analog of (\ref{EC}), this corresponds to the power increase of a blob's volume at the inertial stage.

We see that the account of the behavior of $S_2(r)$ and $u_{eff}(r)$ at large $r$ changes crucially the solution.
The exponential decay of the  correlator is only an intermediate asymptotics, after the time $\sim \lambda_1^{-1} \ln (r_{\eta}/r_d)$ the decay becomes slower.

\subsection{Magnetic field}
The evolution of magnetic field correlator is governed by the Kazantsev equation (\ref{KK}),
$$
\frac{\partial G}{\partial t} =
2 \left( S''+\frac 4r S' \right) G   + \left( 8\frac{S(r)}r + 2S' \right) G' +      2S(r) G'' \ ,
$$
To determine the maximal increment of growth, we add the term $-\gamma G$ in the right-hand side. The effective potential takes the form
$$
2 u_{eff}^{m}= \frac{\gamma}{2S(r)} - \frac 12 \frac{d^2}{dr^2} \ln S(r) - \frac 2r \frac{d}{dr} \ln S(r) -
\frac 34 \left( \frac d{dr} \ln S(r) \right) ^2 +\frac 2{r^2}
$$
In the {\bf Batchelor regime } $r\le r_{\eta}$, $S_2 \propto r^2$  it  can be simplified as
$$
2u_{eff}^{m} =\left(  \frac{3(\gamma - 5D/2)}{2D} - \frac 14 \right) / r^2 \ ,
$$
and the maximal increment is
$$
\gamma_B = \frac 52 D
$$

Unlike the case of scalar field, the magnetic field correlator does not change its behavior significantly after proceeding to the {\bf inertial stage}. The account of  the inertial range asymptotics,
$S_2 \sim r^{2-\epsilon}$ as $r \gg r_{\eta}$ (corresponding effective potential $2u_{eff} \sim -\gamma/ r^{2-\epsilon} \ , \ r \gg r_{\eta}$) changes the increment of growth only slightly:
$$
\gamma_{in} = \gamma_B - O(r_d/r_{\eta})
$$
and the pair magnetic field correlator continues to increase exponentially with the exponent close to that obtained in the  frame of Batchelor approximation.

\vspace{0.6cm} \noindent
{ DATA AVAILABILITY }

The data that supports the findings of this study are available within the article.

\end{document}